# Efficient and Robust *p*-type Transistor based on Ultra-wide-bandgap Semiconductor

Kaijian Xing,[1,2†] Zherui Yang,[3†] Weiyao Zhao,[4] Yuefeng Yin,[4] Huiping Han,[5] Shanhu Wang,[5] Shifan Wang,[6] James Bullock,[6] Alastair Stacey,[7,8] James A. Belcourt,[7] Sergey Rubanov,[9] Hang Yin,[10] David A. Broadway,[7] Jean-Philippe Tetienne,[7] Xinmao Yin,[11] Liang Wu,[5] Dong-Chen Qi,[12*] Michael S. Fuhrer,[2*] Qingdong Ou[1,13*] Xiao Renshaw Wang[3,14*]

1 Macau University of Science and Technology Zhuhai MUST Science and Technology Research Institute, Zhuhai, 519031, China

2 School of Physics and Astronomy, Monash University, Clayton, Victoria 3800, Australia

3 Division of Physics and Applied Physics, School of Physical and Mathematical Sciences, Nanyang Technological University, Singapore, 637371, Singapore.

4 Department of Materials Science & Engineering, Monash University, Clayton, Victoria 3800, Australia

5 Faculty of Materials Science and Engineering, Kunming University of Science and Technology, Kunming, 650093, Yunnan, China

6 Department of Electrical and Electronic Engineering, The University of Melbourne, Victoria 3010, Australia

7 School of Science, RMIT University, Melbourne, Victoria 3000, Australia

8 Princeton Plasma Physics Laboratory, 100 Stellarator Road, Princeton, New Jersey 08540, USA

9 Bio21 Institute, The University of Melbourne, Melbourne, Victoria 3010, Australia

10 Research School of Chemistry, Australian National University, ACT 2601, Australia

11 Shanghai Key Laboratory of High Temperature Superconductors, Department of Physics, Shanghai University, Shanghai 200444, China

12 Centre for Materials Science, School of Chemistry and Physics, Queensland University of Technology, Brisbane, Queensland 4001, Australia

13 Macao Institute of Materials Science and Engineering (MIMSE), Faculty of Innovation Engineering, Macau University of Science and Technology, Taipa, Macao 999078, China
14 School of Electrical and Electronic Engineering, Nanyang Technological University, Singapore, 637371, Singapore.

dongchen.qi@qut.edu.au
michael.fuhrer@monash.edu
qdou@must.edu.mo
renshaw@ntu.edu.sg

[†]These authors contributed equally to this work.

**Abstract**
The *p*-type transistor is an indispensable component of semiconductor technology, enabling complementary operation with *n*-channel transistors for computation, storage, and communication. Achieving both high robustness and high efficiency is highly desirable but challenging for *p*-type transistors due to limited semiconductors with reliable hole transport and their high activation energies. Here, we achieved a robust yet efficient *p*-type transistor by heterogeneously integrating an ultra-wide-bandgap semiconductor and a high-$\kappa$ dielectric layer through van der Waals integration. The *p*-type transistor employs a two-dimensional hole channel on hydrogenated diamond (bandgap 5.6 eV) combined with a high-$\kappa$ (30.5) $SrTiO_3$ perovskite membrane. At room temperature, the transistor exhibits stable operation with a high on-current (~200 mA/mm), low subthreshold swing (70 mV/dec), high hole mobility (566 cm²/Vs to 572 cm²/Vs) and high on-off ratio (~$10^9$). Furthermore, tuning annealing temperature allows operation in either enhancement or depletion mode. The robust *p*-type transistor with high efficiency holds great potential for future power electronics, UV optoelectronics, and harsh-environment electronic applications.

Transistors, classified into two fundamental types—*n*-type and *p*-type—are essential to modern electronics, playing a pivotal role in applications ranging from computing to memory to communications. As electronic technologies progress, the demand for transistors that simultaneously deliver high efficiency and robustness has become critical for reliable operation in advanced fields such as power electronics, optoelectronic devices, and harsh-environment electronics.[1-3] To meet these evolving requirements, there has been a significant shift from conventional semiconductors to wide-bandgap (WBG)[4,5] and ultra-wide-bandgap (UWBG) materials,[6,7] driven by their superior tolerance to high voltage, current, and thermal stress, as well as their ability[8-10] to operate at elevated voltages, frequencies, and temperatures—key factors enabling robust (handling high driven current) and energy-efficient electronics (low subthreshold swing). A WBG semiconductor has a bandgap in the range of 2-4 eV, which enables higher voltage operation, improved thermal stability, and greater resistance to radiation compared to traditional silicon. UWBG semiconductors, with bandgaps exceeding 4 eV, offer even greater advantages over WBG materials as shown in Fig. 1a, but they also present challenges in maintaining the same tunability and efficiency primary due to their larger bandgap.[6] Addressing these challenges in an UWBG semiconductor such as diamond (bandgap 5.5 eV) would offer a route to surpass the capabilities of conventional transistors.

Although achieving both high efficiency and robustness in transistors is inherently challenging for both *n*-type and *p*-type transistors, *p*-type transistors face even greater difficulties, with successful implementations being extremely rare in WBG materials. This is primarily due to the fundamental material and device challenges in *p*-type transistors, such as the limited availability of *p*-type dopants in WBG materials, their high activation energy, and generally low carrier mobilities intrinsic to the valance band technology.[11-13] First, there are only a few *p*-type UWBG semiconductors, such as diamond and aluminum nitride, due to challenges in achieving stable doping and difficulties in controlling material defects.[14-19] Second, *p*-type materials typically have much lower hole mobility (1 to 30 cm$^2$/Vs) compared to their *n*-type counterparts, hindering their ability to handle high power and operate at high speeds.[12, 20-29] Lastly, *p*-type semiconductors suffer from high activation energies (often exceeding 1 eV), which significantly limit their doping efficiency, leading to a poor subthreshold swing in transistors, especially at high current densities.[20, 24, 30-37] These challenges severely limit the progress of high-performance electronics, underscoring the urgent need to enhance the robustness in *p*-type transistors without compromising energy efficiency.

In this work, we demonstrated *p*-type metal-oxide-semiconductor field-effect transistor (MOSFET) with enhanced robustness and energy efficiency based on UWBG hydrogenated diamond (H-diamond) as a semiconducting channel and a high-$\kappa$ perovskite SrTiO$_3$ (STO) membrane as dielectric layer. We prove that surface acceptor removal by annealing reduces scattering resulting in mobility enhancement and two working modes (depletion/enhancement mode). When the transistor works in enhancement mode, high hole mobilities (two-terminal mobility, $\mu_{2T}$ = 565.5 cm$^2$/Vs; four-terminal mobility, $\mu_{4T}$ = 551.7 cm$^2$/Vs) were achieved due to the low-scattering interface between STO and diamond. The device also presents high on-off ratio (~ 10$^9$), low subthreshold swing (~ 70 mV/dec) and high-ON current (~ 200 mA/mm). Moreover, while maintaining the abovementioned efficiency and mobility, a high ON gate length normalized current density of ~4000 um·A/mm is shown to demonstrate the robustness of the *p*-type MOSFET.

**Results and Discussion**

*Interface Engineering*

Fig. 1b to Fig. 1d outline the conceptual framework of achieving high-performance WBG semiconductor transistors by integrating high-$\kappa$ STO with UWBG H-diamond. The STO membrane was physically transferred onto the H-diamond surface to achieve the van der Waals (vdW) integration (Fig. 1b). Notably, the transferred STO membranes can be mechanically peeled off from the H-diamond surface using a dry transfer method (c.f. supporting video), indicating the vdW nature of the STO/H-diamond interface.[38] This approach presents notable advantages over conventional methods for dielectric layer fabrication, including thermal evaporation, pulsed laser deposition (PLD), and atomic layer deposition (ALD), as they are prone to introducing defects, trap states, and scattering centers, ultimately compromising interface quality and device performance, which greatly hampers the development of electronic applications based on surface-conducting diamonds. In addition, the high energy process can strip off the surface-termination hydrogen, resulting in the degradation of the device performance. The low-energy nature of low-temperature physical transfer is particularly advantageous for surface-conducting diamond devices, as it preserves the hydrogen termination which is essential for the formation of a hole accumulation layer at the diamond surface,[39] while minimizing defect and trap formation compared to conventional techniques. Additionally, vdW integrating oxides materials (2D oxides or three-dimensional oxide membranes) can affect the interface qualities[40-42] and heterostructure performances.[43] The interface quality was assessed via cross-sectional transmission electron microscopy (TEM) and energy-dispersive X-ray spectroscopy (EDS). As shown in Fig. 1c and Fig. S1, the STO/diamond interface exhibits sharp boundary without interfacial reaction layer. No evidence of lattice intermixing, dislocations, or chemically bonded interfacial phases is detected within the resolution of the measurements, consistent with a weakly interacting interface. The interface further shows minimal damage, interdiffusion, or contamination (including photoresist residues) confirming its structural integrity. This high-quality interface achieved by physical transfer with little evidence of damage is pivotal for enabling high-performance *p*-type transistor operation, as listed in Fig. 1d.

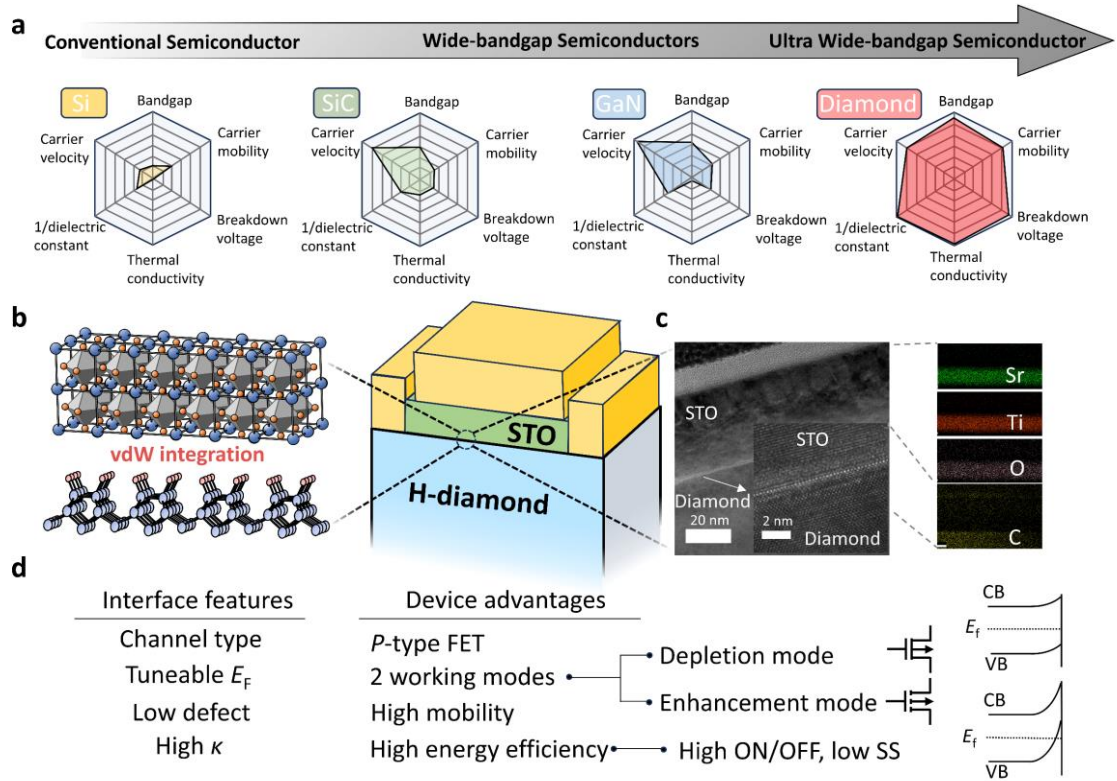

**Fig. 1. Integration of STO membranes on H-diamond surface by physical transfer towards high performance FETs.** (**a**) Key electronic parameters for conventional semiconductor (Si), wide-bandgap semiconductor (SiC, GaN) and ultra-wide-bandgap semiconductor (diamond). (**b**) Schematic of STO/H-diamond FET. (**c**) Cross-sectional TEM images (inset shows a magnified view of the highlighted interfacial region) and EDS mapping of STO/H-diamond interface. (**d**) Details the advantages of the device presented in this work.

Enhancement and depletion modes represent the two primary working modes of transistors used in electronic devices. For example, enhancement-mode FETs are typically utilized as switching elements, while depletion-mode FETs often serve as load resistors in logic circuits. Here, we first characterized the working mode tunability as a function of pre-anneal temperatures before the transfer of free-standing STO membranes onto H-diamond. It is known that annealing H-diamonds can reduce the diamond-surface conductivity due to the desorption of atmospheric adsorbates which otherwise act as surface acceptors giving rise to the 2D hole layer on H-diamond.[39, 44] Surface acceptors can be mostly removed while keeping hydrogen termination intact by annealing at moderate temperature (300°C ~ 400°C) under ultra-high vacuum condition.[45, 46] In this work, we performed the annealing in a glovebox filled with Nitrogen ($O_2$ < 0.1 ppm, $H_2O$ < 0.1 ppm) to minimize the possible oxidation of the diamond surface. The initial surface sheet resistance of H-diamond was measured as 11 kΩ/□ by a four-point probe setup and the sheet resistance of diamond surface experienced an increase with increasing annealing temperature as shown in Fig. S2a. The diamond surface lost its conductivity completely (beyond the equipment measurement capability) when annealing is above 110°C. The diamond surface is able to recover its sheet resistance to 10 kΩ/□ after re-exposing to atmosphere for few minutes, indicating that the surface hydrogen termination is intact during this annealing process in an $N_2$ environment.

Then we performed *I-V* characterization on a two-terminal device at five states: (1) before STO transfer; (2) after glovebox annealing at 110 °C; (3) after STO capping of the entire channel in the glovebox; (4) after peeling off STO in the glovebox; and (5) after exposure to ambient atmosphere. As shown in Fig. S2b, state (1) and (5) exhibit similar two-terminal conductance (almost identical *I-V* curves), demonstrating the hydrogen-terminated diamond surface remains intact throughout the STO capping and peeling processes. The unmeasurable two-terminal conductance from states (2) to (4) indicates that the physically transferred STO has a negligible influence on the H-diamond surface, consistent with a weak interfacial interaction. This observation agrees well with the DFT results in Fig. 2a, which show negligible interfacial charge transfer at the STO/H-diamond interface. Furthermore, for a two-terminal device capped with STO after glovebox annealing and subsequently exposed to ambient conditions, the *I–V* curve remains unmeasurable, demonstrating that STO capping effectively suppresses the recovery of diamond surface conductivity upon air exposure. The wide tunability of the surface-conductivity of diamond allows us to change the working mode of diamond-based FETs by controlling the pre-annealing temperatures before STO landing. As shown in Fig. 2b and Fig. 2c, by adjusting the annealing temperature, the threshold voltages ($V_{th}$) of STO/H-diamond FETs shift from 3.7 V to –2.6 V, corresponding to a transition from depletion mode to enhancement mode. We propose that the tunability in working modes can be explained by weakened band bending at H-diamond surface as a result of a shift of Fermi level due to the desorption of surface acceptors with annealing. As shown in the insets of Fig. 2b, mild annealing at 60 °C is only able to partially remove the surface acceptors, indicating the slight reduction of the band bending at H-diamond surface, and eventually leading to an on state at zero gate voltage (depletion mode). With pre-annealing temperature increasing from 60 °C to 110 °C, the upward band bending of the valance band was weakened as surface acceptors were desorbed. As a result, the surface Fermi level shifts back to the band gap and the device works in enhancement mode with a normally-off state. It should be noted that the normally-off state is not contributed by the gate metal depletion effect because the work function of palladium (Pd) (approximately 5.6 eV) is similar to the ionization potential of the H-diamond surface. Following this, we further studied the relationship between two-terminal field-effect mobilities and the annealing temperatures. As depicted in Fig. 2c, the two-terminal field-effect mobilities are highly dependent on the annealing temperature before STO landing. The field-effect mobility for an annealed device at 110 °C is 565.5 cm$^2$/Vs, which is significantly higher than the annealed device at 60 °C (~230 cm$^2$/Vs). The significant increase in mobility is mainly attributed to reduction of the surface charged impurities, as the Coulomb scattering, caused by charged surface acceptors, is the dominant mechanism that limits diamond-based FET mobilities.[47] Further improvements in STO (surface quality), H-diamond surface roughness and optimization of fabrication processes may push the hole mobility to the theoretical limitation in this system,[47] which could be the focus of future independent studies. It should be noted that even with the depletion mode transistor, the field-effect mobilities are still superior to the reported values in the literature,[48-57] which indicates that the vdW dielectric integration offers a more effective way to form a high-quality interface between the dielectric and diamond, over conventional methods such as ALD or thermal evaporation.

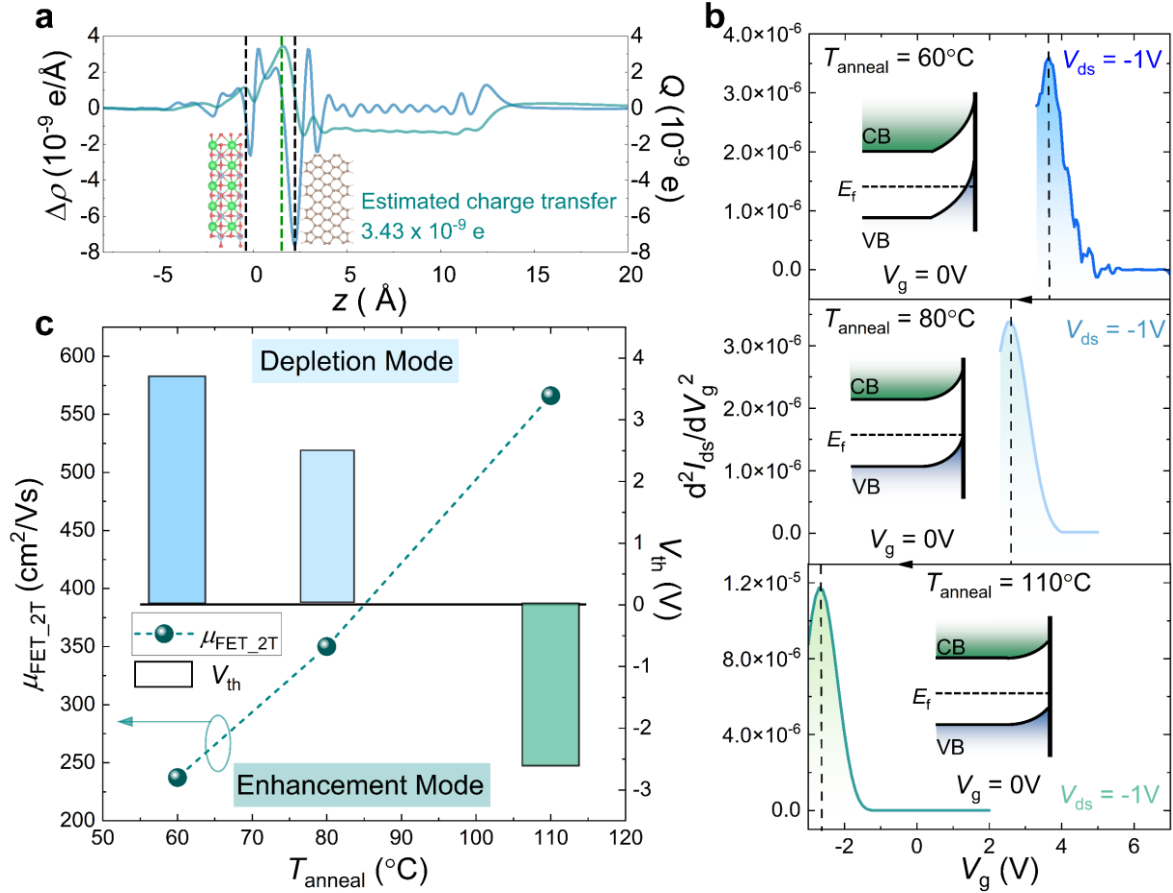

**Fig. 2. Tunability of working mode for STO/H-diamond FETs.** (**a**) The planar-averaged charge density difference for the H-diamond/STO interface. (**b**) $d^2I_{ds}/dV_g^2$ as a function of gate bias for STO/H-diamond FETs with different annealing temperatures. The threshold voltage is estimated to be 3.6 V for $T_{anneal} = 60°C$, 2.5 V for $T_{anneal} = 80°C$ and –2.6 V for $T_{anneal} = 110°C$. The insets in each subfigure qualitatively illustrate the band-bending states at different annealing temperatures. (**c**) Threshold voltage and field-effect mobilities as a function of annealing temperatures.

*MOSFET Room-temperature Characterization*

Next, we evaluated the room temperature electrical properties of four terminal STO/H-diamond FETs with the 110 °C pre-annealing. Fig. 3a shows both two-terminal and four-terminal transfer curves (conduction as a function of gate bias) with a constant $V_{ds}$ of –1V and inset of Fig. 3a shows the source-drain current as a function of gate bias, suggesting that the device shows a typical *p*-type enhanced mode behaviour. The $V_{th}$ is approximate –2.6V which has been discussed above. The fitted two-terminal FE mobility is 566 cm$^2$/Vs and the four-terminal FE mobility is 552 cm$^2$/Vs, which are determined by $\mu_{2T} = \left(\frac{1}{C_{ox} \cdot V_{ds}}\right)\left(\frac{L}{W}\right)\frac{\partial I_{ds}}{\partial V_{gs}}$ for two-terminal measurement and $\mu_{4T} = \frac{1}{C_{ox}} \cdot \frac{D}{W} \cdot \frac{\partial (I_{ds}/V_{4T})}{\partial V_{gs}}$ for four-terminal measurement, respectively. In these equations, $C_{ox}$ is the capacitance of STO, $L$ (~30 μm) and $W$ (~2 μm) represent the channel length and width and $D$ (~20 μm) is the distance between two inner electrodes. These two values are nearly identical which is due to the excellent Pd Ohmic contacts at room temperature.[58] The on-off ratio is ~ $10^9$ at $V_g$ = –10 V with the $V_{ds}$ of –6 V as shown in Fig. 3b. The extracted subthreshold swing is approximately 70 mV/dec, which is the lowest reported value and close to the thermal dynamic limit (60 mV/dec). The interface trap density, $D_{it}$, is calculated as $6.5 \times 10^{11}$ cm$^{-2}$eV$^{-1}$ (by Eq. S1 in the Supporting Information), which is lower than the control Al$_2$O$_3$/H-diamond MOSFET ($4.2 \times 10^{12}$ cm$^{-2}$eV$^{-1}$) and most of the reported H-diamond based MOSFETs as listed in Table S1. The low $D_{it}$ indicates a high-quality interface between STO membrane and H-diamond surface, which is consistent with the high-mobility behavior. The output curves show the excellent performance of the STO/H-diamond FET, both Ohmic and saturation regions can be clearly observed as shown in Fig. 3d. We noticed a small saturation current reduction (which is less than 2%) at high $V_{ds}$, and we hypothesize that it is likely due to the lateral gate effect. The maximum normalized current is determined to be ~ 200 mA/mm with a $V_g$ of –16 V at $V_{ds}$ of –42 V (Fig. 3e). The survival of this device at such a high $V_{ds}$ also shows the robust nature of H-diamond based electronics compared with other systems such as thin metal oxide materials and it should be noted that the $V_{ds}$ is limited by the output of the source meter. Despite the long channel length of this device (30 µm), this on-current is exceptionally high among the reported H-diamond based transistors. This four-terminal device was further evaluated after two months in atmosphere without passivation protection to assess its stability over time. As shown in Fig. S3a, the device retained stable *p*-type transistor behavior after two months. The two-terminal mobility decreased by ~9%, from 565 to 510 cm²/Vs; the subthreshold swing increased from 70 to 88 mV/dec; and the threshold voltage remained nearly unchanged (Fig. S3b). These variations can be attributed to minor changes at the heterointerface between STO and the H-diamond surface. Overall, the STO/H-diamond MOSFET exhibits excellent long-term stability.

To quantitatively assess device-to-device variation and reproducibility, another four STO/H-diamond MOSFETs with identical fabrication process were characterized (Fig. S4). All devices exhibit comparable transfer and output characteristics, as well as similar key performance metrics, including field-effect mobility, subthreshold swing and threshold voltage, indicating high uniformity and reproducibility. Statistical analysis of the extracted key metrics is summarized in Fig. S5. The two-terminal field-effect mobility shows a mean value of approximately $530 \pm 30$ cm$^2$/Vs, corresponding to a relative standard deviation of ~6%, reflecting limited variability in carrier transport across devices (Fig. S5a). The subthreshold swing is tightly distributed with an average of $72 \pm 4$ mV/dec (Fig. S5b), suggesting a uniform STO/H-diamond interface quality and reproducible gate dielectric performance. In addition, the threshold voltage exhibits a mean value of $-2.2 \pm 0.4$ V (Fig. S5c), indicating minimal dispersion in fixed charge density and electrostatic doping conditions among devices. The small standard deviations observed for all key parameters demonstrate excellent device-to-device uniformity and highlight the reliability of vdW integration of high-$\kappa$ STO membranes on H-diamond for realizing high-performance MOSFETs.

*Dielectric Constant Characterization*

We also investigate the dielectric performance of the STO membrane by characterizing the gate leakage current and the capacitance-voltage response. As shown in Fig. S6, the leakage current density, $J_{leakage}$, is a few orders of magnitude smaller than the low power limit ($10^{-2}$ A/cm$^2$) for complementary-metal-oxide-semiconductor devices. It demonstrates that the transferred high-quality STO membrane can sufficiently suppress the gate leakage. Capacitance-Voltage measurements taken at 10 kHz are shown in Fig. 3f and clearly reveal the accumulation and depletion states of the STO/H-diamond FET. The onset transition happened at $V_g$ of –2 V, which is consistent with the enhancement mode. The insets of Fig. 3f illustrate the band alignment states of the STO/H-diamond FET at different $V_g$. When the gate bias moves to the negative side, strong band bending occurs at diamond surface due to the interfacial electrostatic field, resulting in hole accumulation at the diamond surface. On the other hand, diamond's valance band becomes flat when the gate bias moves to the positive side, resulting in hole depletion. We further calculated the dielectric constant of the STO membrane based on $\kappa = \frac{C \times A}{d} / \varepsilon_0$ yielding a value of 30.7, where $C$ is the measured capacitance of dielectric layer, A is the effective area of the capacitor, d is the thickness of the STO membrane, and $\varepsilon_0$ is the vacuum permittivity. This value is comparable with the value (30.5) extracted from metal-insulator-metal device measurement as shown in Fig. S7. Here, we used $h$BN flakes as reference samples to confirm the accuracy of the capacitance measurements. The STO dielectric membrane clearly exhibits superior dielectric performance in comparison to other dielectric materials used in diamond (solid triangles) and non-diamond devices (green bars) (Fig. 3g), with its dielectric constant significantly beyond the high-$\kappa$ benchmark (blue dashed line), suggesting the high-quality interface formed by vdW integration. Furthermore, the 60 nm STO membrane exhibits a low effective oxide thickness (EOT) of 7.6 nm according to Fig. S7e, which is significantly lower than that of the reference vdW dielectric $h$BN flakes (~10 nm), as shown in Fig. S7e. The combination of low leakage current and reduced EOT ensures the high performance of diamond-based electronic devices.

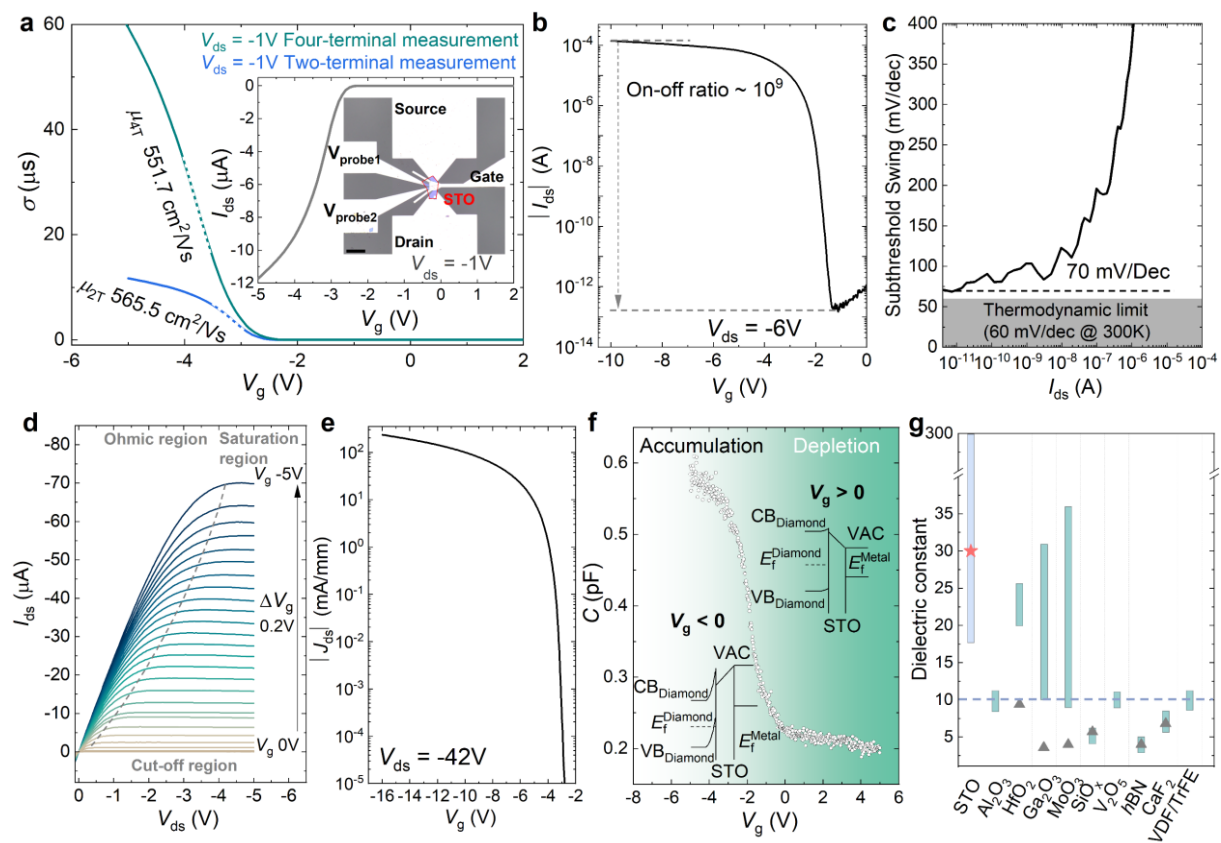

**Fig. 3. Transport characterization of high-performance STO/H-diamond FET at room temperature.** (**a**) Two-terminal and four-terminal transfer characterization of STO/H-diamond FET at $V_{ds} = -1$ V. Inset shows the two-terminal source-drain current as a function of gate bias and an optical image of the measured four-terminal STO/H-diamond FET. The scale

bar represents 50 μm. (**b**) is the transfer curves at $V_{ds}$ = -6V (**c**) Subthreshold swing as a function of source-drain current. (**d**) Output curves with gate bias varying from 0V to –5V with a stop of 0.2V. (**e**) Current density as a function of gate bias at high $V_{ds}$ (–42V). (**f**) *C-V* measurements at 10 kHz. Inset panels depict the band diagram for depletion region ($V_g > 0$) and accumulation region ($V_g < 0$). (**g**) shows the κ values for different dielectric materials, bars represent measured κ values region in other systems, solid star and triangles represent the measured κ values of the dielectric in diamond-based transistors. The blue dash line represents the benchmark of high κ.

*Temperature Dependent Characterization*

The 2D hole gas accumulated at the H-diamond surface has also garnered significant attention due to the observation of various quantum transport phenomena at cryogenic temperatures, including weak localization/weak antilocalization,[59, 60] Shubnikov–de Haas (SdH) oscillations,[61, 62] and giant spin-orbit interaction.[63, 64] Consequently, achieving tunable surface conductivity at low temperatures is of considerable importance. However, conventional gating strategies are often ineffective at low temperatures due to the presence of defects or interfacial traps, which may lead to carrier freeze-out[65] at cryogenic temperatures. As shown in Fig. S8, the transfer curves for the STO/H-diamond MOSFET experienced a moderate reduction with decreasing temperature, while the transfer curves for the control Al$_2$O$_3$/H-diamond MOSFET became unmeasurable at low temperatures. In addition, as shown in Fig. 4a, the channel current of a control device, Al$_2$O$_3$/H-diamond FET is strongly dependent on temperature, and exhibits a significant reduction with decreasing temperature, ultimately reaching the off-state at approximately 40 K when $V_g$ is –5 V and $V_{ds}$ is –2 V. In contrast, the channel current of the STO/H-diamond FET experiences only a slight reduction with decreasing temperature (less than an order of magnitude) and eventually becomes almost independent of temperature below 20 K. The observed decrease in source-drain current can be attributed to both contact effects and hole-hole interactions.[59] This behaviour indicates an excellent interfacial condition with minimal defect traps and scattering centers. Fig. 4b and 4c illustrate the dependence of channel sheet resistance on both temperature and gate bias, as determined through four-terminal measurements. At 300 K, the sheet resistance transitions from an insulating state to approximately 8 kΩ/□ under a small gate bias of –5 V. As the temperature decreases from 300 K to 10 K, the sheet resistance of the active channel increases from approximately 8 kΩ/□ to 56 kΩ/□ at the same gate bias. After normalization with respect to $V_g - V_{th}$, the sheet resistance increases from ~ 8 kΩ/□ to ~ 40 kΩ/□. This increase is attributed to hole-hole interactions within this two-dimensional hole gas system, as the temperature dependence follows a logarithmic rather than an exponential trend with temperature, as shown in Fig. S9, which is characteristic of a two-dimensional Fermi liquid.[60] As shown in Fig. 4d, although output measurements suggest the presence of Schottky contacts between Pd electrodes and H-diamond surface at the base temperature (1.8K), charge carriers can still tunnel through the Schottky barrier and be injected into the channel. The contact performance could be further enhanced through a charge-transfer electrode strategy.[66] These findings suggest that the diamond surface can be modulated from an insulating towards a metallic state through the STO gating architecture, which is crucial for investigating emergent quantum phenomena in surface-conducting diamond systems.

High-temperature operation of the STO/H-diamond MOSFET was also evaluated, as shown in Fig. S10. The device remained functional up to 400 K (Fig. S10a). The threshold voltage shifted from –2.4 to –2.8 V (Fig. S10b), and the on-current at $V_g$ = –4 V decreased from 3.5 × $10^{-5}$ to $2 \times 10^{-5}$ A (Fig. S10c). In addition, the two-terminal mobility exhibited less than a 10% reduction (Fig. S10d). Overall, the device demonstrates excellent high-temperature stability up to 400 K (the temperature range being limited by the measurement equipment).

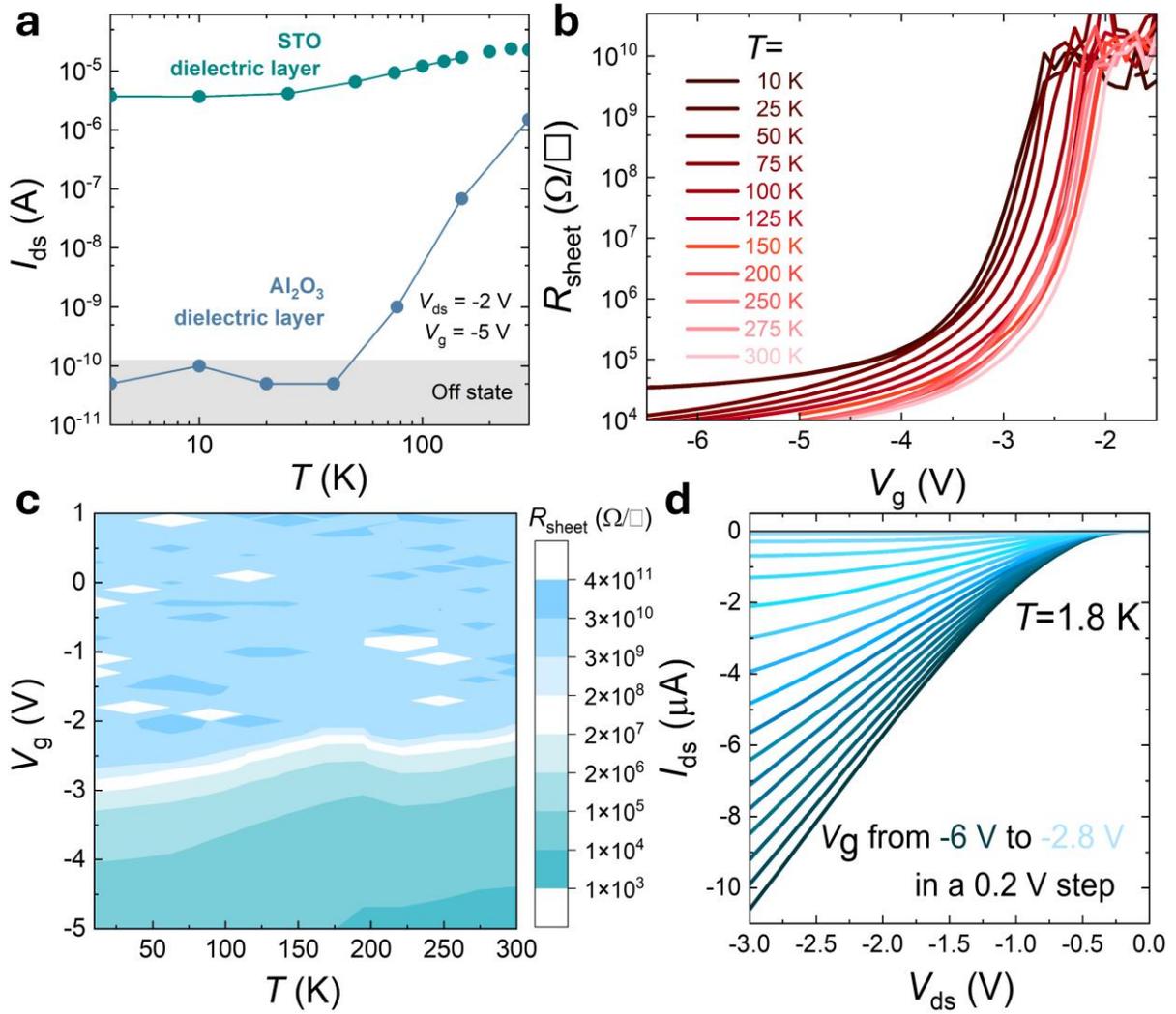

**Fig. 4. Electrical performance of STO/H-diamond FET as a function of temperature.** (**a**) on current of diamond-based FET as a function of temperature with different dielectric layers. Red line represents the STO/H-diamond FET, the blue line represents the $Al_2O_3$/H-diamond FET. (**b**) and (**c**) depict the channel sheet resistivity as a function of gate bias at different temperatures. (**d**) shows the output curves at base temperature (1.8K).

As summarized in Table S2, the STO/H-diamond MOSFET (Fig. 3) demonstrates superior performance to the control $Al_2O_3$/H-diamond MOSFET (Fig. S11). We then benchmark the room-temperature performance of the STO/H-diamond FET reported here against a broad range of *p*-type WBG transistors from the literature (Fig. 5). As shown by Fig. 5a, the STO/H-diamond FET shows one of the highest field-effect mobilities among the H-diamond-based FETs which, as a result of the unique benefits from the vdW integration process that minimizes the formation of defects and traps at the interface. In addition, the highest gate-length normalized current density suggests that STO/H-diamond FET shows great potential in high power device applications. As depicted in Fig. 5b, benefiting from the high *κ* nature of the STO membrane, our diamond FET also achieves high on-off ratio and low subthreshold swing. Additionally, the STO/H-diamond FET is superior to other *p*-type transistors based on emerging electronic materials, such as two-dimensional semiconductors (Table S3) and other wide-band semiconductors (SiC, GaN: green dots in Fig. 5a and 5b). Therefore, exploiting vdW integration of dielectric layers to H-diamond surfaces presents a promising pathway to overcome several limitations associated with device operation compared with conventional methods for dielectric formation. Fig. 5c summarizes prior studies that have employed vdW integration (physical transfer process) strategies to form vdW (*h*BN) or no-vdW (amorphous $Ga_2O_3$) dielectric layers on H-diamonds, providing a comparative analysis of device performance across multiple parameters. Notably, the integration of STO with H-diamond exhibits a substantial improvement in device performance compared to the use of amorphous $Ga_2O_3$. This enhancement is attributed to the fact that the amorphous $Ga_2O_3$ layer, formed via the liquid metal method, introduces numerous defect traps both in the dielectric layer and at the interface, thereby leading to significant device performance degradation. In contrast, the single-crystal STO membrane establishes a high-quality interface with H-diamond surface, effectively mitigating these detrimental effects and enhancing overall performance. Furthermore, the STO/H-diamond device demonstrates a superior dielectric constant and subthreshold swing relative to *h*BN (low-$\kappa$ dielectric)/H-diamond devices. Therefore, vdW integrating high $\kappa$ dielectric (STO) on H-diamond surface paves the way for enhancing the performance of diamond-based electronics.

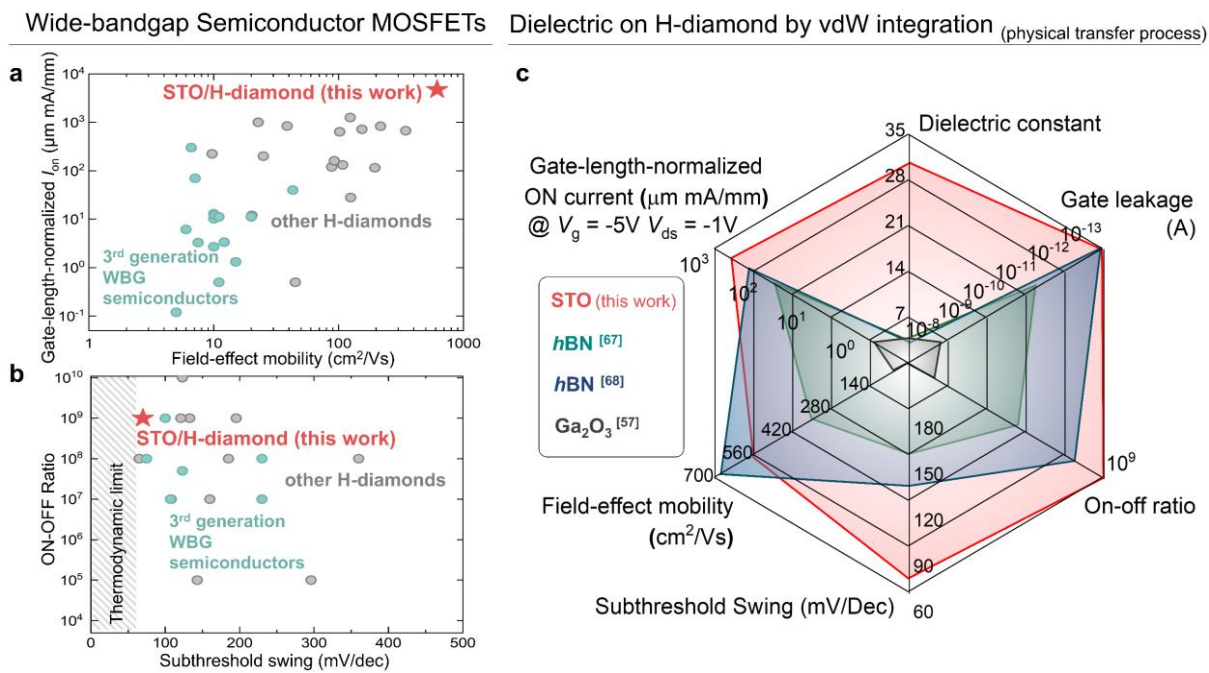

**Fig. 5. Comparison of STO/H-Diamond FET performance against other diamond-based FETs and *p*-type WBG competitors.** (**a**) Shows normalized on current and field-effect mobilities, and (**b**) Shows on-off ratio and subthreshold swing of our present device (red star) and other *p*-type WBG and H-diamond FETs. The grey dots represent other diamond-based FETs. The green dots represent the FETs based on other wide-bandgap semiconductors, such as GaN or SiC. Data in (**a**) and (**b**) are summarized in Table S3. (**c**) compares the diamond-based FETs with different dielectrics by physical transfer process in terms of normalized on current, extracted dielectric constant in devices, gate leakage, field-effect mobility and on-off ratio.[57, 67, 68]

**Conclusion**

In this study, we concurrently improved energy efficiency and robustness of the *p*-type transistor by integrating the UWBG of H-diamond with a high-$\kappa$ dielectric membrane with high-quality interfaces. By optimizing the annealing temperature prior to STO lamination, we achieve controlled engineering of two operational modes—depletion mode and enhancement mode. Notably, the interface engineering on the surface acceptors significantly enhances hole mobility by reducing scattering centers. The enhancement-mode device exhibits high hole mobility (566 cm²/Vs to 572 cm²/Vs), excellent on/off current ratio (~$10^9$), low subthreshold swing (70 mV/dec), and high on-current (~200 A/mm), marking a significant advance in *p*-type MOSFET technology. Recent progress in wafer-scale dielectric layer transfer[69-71] provides a possible path for the practical realization of large-scale STO/H-diamond MOSFET technologies. This work highlights a promising route toward high-performance, thermally stable, and reliable electronics capable of operating under extreme conditions, and opens new opportunities for UWBG device platforms.

**Experimental Section**

*Materials*

Hydrogen termination of diamonds: The purchased (100) diamond substrates were firstly polished by scaif wheel to optimize the surface roughness. Then a Seki 6500 diamond deposition reactor with a 2.4 GHz microwave plasma assisted CVD chamber was employed to process the hydrogen termination. We follow our previous hydrogen termination recipe[72] to minimize the etching pits as a result of a smooth diamond surface down to nano-meter regime.

STO membrane preparation (Fig. S12a): Before STO/$Sr_3Al_2O_6$ (SAO) deposition, STO substrates were treated with mixture of concentrated nitric and hydrochloric acids.[73] Then STO/SAO bilayer films were sequentially deposited by pulsed laser deposition (PLD) on $TiO_2$-terminated STO (001) substrates utilizing sintered SAO and STO ceramic targets.[73] During growth, the materials were vaporized by an excimer laser ($\lambda$=248 nm) operating at a repetition rate of 2 Hz and a fluence of 1.8 J/cm², meanwhile the STO substrates were maintained at 760 °C. Oxygen partial pressure for SAO and STO growth was $1 \times 10^{-2}$ and $1 \times 10^{-2}$ torr, respectively. After deposition, all samples were cooled down to room temperature at the rate of 10 °C min$^{-1}$ under $PO_2$ ambient of $1 \times 10^{-2}$ torr. The surface of STO thin film was then covered with a polydimethylsiloxane (PDMS) sheet. The stack of STO-PDMS sheet was then floated in a deionized water bath overnight to separate the SAO layer from the STO film. By repetitively laminating and peeling the PDMS sheets, STO flakes of micrometer-size were extracted from the original stack, and then an appropriate one was selected for the following diamond-based device fabrication.

*Device Fabrication*

STO/H-diamond FET fabrication (Fig. S12b): Firstly, a (100) diamond substrate was hydrogen terminated in a Seki6500 diamond deposition reactor, consisting of a 2.4 GHz microwave plasma-assisted chemical vapor deposition. (Fig. S12b ①) The hydrogenated active channels were firstly defined and isolated by the standard photolithography and oxygen plasma (Fig. S12b ②). Then the Pd contacts were fabricated by the standard photolithography, e-beam evaporation and lift-off process to achieve ohmic contacts[58] as shown in Fig. S12b ③. Following diamond pre-anneal process in a glove box filled with Nitrogen, the selected STO flakes were *in-situ* picked up and laminate on the hydrogenated channels by dry transfer technology (propylene carbonate method) (Fig. S12b ④). For the control $Al_2O_3$/H-diamond FET, ALD was employed to form the $Al_2O_3$ layer on hydrogenated diamond surface. The formation temperature was controlled at 150 ºC. Afterwards, Pd electrodes were fabricated by photolithography, e-beam evaporation and lift-off processes to achieve the top gate electrode (Fig. S12b ⑤).

STO Capacitor fabrication: Bottom electrodes (Ti/Au) were first fabricated on sapphire substrates by a standard photolithography and lift-off process. Then the STO membrane was picked up and transferred on to the bottom electrode by a dry transfer technique (PPC method). Lastly, the top electrodes were achieved by the standard lithography technique and lift off process to achieve the MIM architecture. We also fabricated *h*BN capacitors following the same fabrication process for calibration and comparison.

*Experimental Measurements*

Surface conductivity characterization: A custom designed four-point-probe stage integrated with a Keithley source meter (2450) was employed to measure the sheet resistance of hydrogenated diamond surface at different temperatures in glovebox environments ($N_2$ filled).

FET characterization: The Oxford TeslatronPT assisted with two Keithley 2400 source meters and a multimeter (Agilent 34401A) were employed to perform the DC transport measurements at different temperatures. One Keithley was used for applying the gate voltage, the other one was used to apply the source-drain voltage and monitor the source drain current while the multimeter measured the four-probe voltage. A Keysight B1500 semiconductor analyser assisted with a probe station was employed to characterize the FET performance in terms of DC transport and *C-V* measurement in atmosphere. Probe station (Lakeshore) integrated with Keithley 2400 source meters was employed to operate the high-temperature characterization for STO/H-diamond MOSFETs from 300 to 400 K.

STEM and EDS Mapping: Cross-sectional TEM and energy dispersive X-ray spectroscopy (EDS) elemental mapping was performed using FEI Tecnai TEM operated at 200 keV in TEM mode. To reduce the ion damage introduced into TEM lamella during preparation, focus ion beam (FIB) milling (FEI Nova Nanolab 200) at 5 keV was utilized at the final stage.

*DFT Calculation Details*

We used density functional theory (DFT) calculations as implemented in the Vienna Ab initio Simulation Package (VASP) to calculate the charge transfer between STO and H-diamond.[74] The Perdew-Burke-Ernzerhof (PBE) form of the generalized gradient approximation (GGA) was used to describe electron exchange and correlation.[75] The kinetic energy cut-off for the plane-wave basis was set to 400 eV. We used 5×*5*×1 supercell of STO and 4×*4*×1 supercell of H-Diamond to model the interface. The lattice mismatch between the two materials is 3.1%. We used a 3×*3*×1 k-point mesh for sampling the Brillouin zone. A 20 Angstrom vacuum was added to prevent interactions between periodic images. To obtain the relaxed interface between the metal and the substrate, we used the lattice of the substrate, allowing all atoms to relax except for the bottom two metallic layers. All forces were relaxed to less than 0.01 eV/Å. The DFT-D3 method was used to account for the van der Waals interactions.[76]

**Supporting Information**

Interface trap density calculation, cross-section TEM images for STO/H-diamond interface, sheet resistance as a function of annealing temperature and two-terminal conductance characterization at different states, STO/H-diamond MOSFET stability over time, device-to-device variation for STO/H-diamond MOSFETs, statistical summary for the presented STO/H-diamond MOSFETs in this work, gate leakage current density as a function of gate bias at different $V_{ds}$, dielectric constant characterization, temperature-dependent transfer curves of STO/H-diamond MOSFET and $Al_2O_3$/H-diamond MOSFET, H-diamond channel conductance as a function of temperature at different gate bias, STO/H-diamond MOSFET high-temperature operation, electrical transport characterization of control $Al_2O_3$/H-diamond MOSFET, device fabrication process, summary of the interface trap density of H-diamond transistors, device performance comparison between the STO/H-diamond MOSFET and the control device, summary of room-temperature device performance for 2D-based and WB-based and diamond-based MOSFETs in Fig. 5.

**Acknowledgments**

D.-C. Q. acknowledge continued support from the Queensland University of Technology (QUT) through the Centre for Materials Science. This work was performed in part at the Melbourne Centre for Nanofabrication (MCN) in the Victorian Node of the Australian National Fabrication Facility (ANFF). Y.Y. acknowledges the assistance of resources from the National Computational Infrastructure (NCI Australia), an NCRIS enabled capability supported by the Australian Government.

K. X. and M. S. F. acknowledge Australian Research Council Discovery Projects grant (DP200101345)
M. S. F. acknowledges Australian Research Council Centre of Excellence in Future Low-Energy Electronics Technologies (FLEET; CE170100039)
D. A. B. acknowledges Australian Research Council Discovery Early Career Researcher Award (Grant No. DE230100192)
L. W. acknowledges Xingdian Talent Support Project of Yunnan Province (Grant No. KKRD202251009)
D. C. Q. acknowledges Australian Research Council Discovery Projects grant (Grant No. DP230101904)
A. S. acknowledges Australian Research Council Linkage Projects grant (Grant No. LP190100528)
J. B. acknowledges Australian Research Council Discovery Early Career Researcher Award (Grant No. DE210101129)
Q. O. acknowledges the Science and Technology Development Fund, Macau SAR (0065/2023/AFJ, 0116/2022/A3), the National Natural Science Foundation of China (52402166), the Natural Science Foundation of Guangdong Province (2025A1515011120) and Australian Research Council Discovery Early Career Researcher Award (Grant No. DE220100154)
X. R. W. acknowledges Singapore Ministry of Education under its Academic Research Fund (AcRF) Tier 1 (Grant No. RG82/23 and RG155/24) and Tier 3 grant (MOE-MOET32023-0003) "Quantum Geometric Advantage" and the National Research Foundation (NRF)

Singapore under its 21st Frontier Competitive Research Programs (Grant No. NRF-F-CRP-2024-0012)

**Author contributions:**

Conceptualization: KX, MSF, DCQ, QO, XRW
Methodology: KX, HH, SW, LW, ZY, XRW, JAB, AS
Investigation: KX, WZ, YY, SW, JB, SR
Visualization: KX, HY, XRW
Supervision: KX, QO, XRW
Writing—original draft: KX, ZY, XRW
Writing—review, editing & discussion: KX, ZY, WZ, YY, HH, SW, SW, JB, AS, SR, HY, DAB, JPT, XY, LW, DCQ, MSF, QO, XRW

**Conflict of Interest**

The authors declare no conflict of interest.

**Data Availability Statement**

The data that support the findings of this study are available from the corresponding author upon reasonable request.